\documentclass[a4paper]{jpconf}

\usepackage{upgreek}
\usepackage{enumerate}
\usepackage{float}
\restylefloat{table}
\usepackage{graphicx}
\usepackage[all]{xy}
\usepackage{amsmath}
\usepackage{amsthm}
\usepackage{amssymb}
\usepackage[usenames,dvipsnames]{xcolor}

\def\cW{\mathcal{W}}
\def\codim{\mathrm{codim}}

\def\alg{\mathrm{alg}}
\def\Spec{\mathrm{Spec}}
\def\Ext{\mathrm{Ext}}
\def\bvarphi{\boldsymbol{\varphi}}
\def\NS{\mathrm{NS}}
\def\HF{\mathrm{HF}}

\newcommand{\Hom}{{\rm Hom}}

\newcommand{\ben}{\be}
\newcommand{\een}{\ee}
\newcommand{\beqan}{\beqa}
\newcommand{\eeqan}{\eeqa}

\newcommand{\im}{\mathrm{im}}
\newcommand{\End}{\mathrm{End}}

\newcommand{\eqdef}{\stackrel{{\rm def.}}{=}}
\newcommand{\cA}{{\cal A}}

\newcommand{\cL}{{\cal L}}
\newcommand{\Z}{\mathbb{Z}}

\newcommand{\str}{{\rm str}}

\def\HDF{\mathrm{HDF}}
\def\Pic{\mathrm{Pic}}
\def\P{\mathbb{P}}

\DeclareFontFamily{U}{rsf}{}
\DeclareFontShape{U}{rsf}{m}{n}{<5> <6> rsfs5 <7> <8> <9> rsfs7 <10-> rsfs10}{}
\DeclareMathAlphabet\Scr{U}{rsf}{m}{n}

\def\C{{\mathbb{C}}}

\def\Z{{\mathbb{Z}}}

\def\rk{{\rm{rk}}}
\def\deg{{\rm{deg}}}

\def\cO{\mathcal{O}}

\def\Ob{\mathrm{Ob}}
\def\O{\mathrm{O}}
\def\Res{\mathrm{Res}}

\newcommand{\id}{\mathrm{id}}
\newcommand{\be}{\begin{equation*}}
\newcommand{\ee}{\end{equation*}}
\newcommand{\beqa}{\begin{eqnarray*}}
\newcommand{\eeqa}{\end{eqnarray*}}
\newcommand{\nn}{\nonumber}

\def\cH{\mathcal{H}}
\def\cT{\mathcal{T}}

\def\HPF{\mathrm{HPF}}
\def\PF{\mathrm{PF}}
\def\Jac{\mathrm{Jac}}
\def\F{\mathrm{F}}
\def\HF{\mathrm{HF}}

\def\i{\mathrm{i}}

\def\MF{\mathrm{MF}}
\def\HMF{\mathrm{HMF}}

\newtheorem{Theorem}{Theorem}[section]

\newtheorem{Proposition}[Theorem]{Proposition}

\newtheorem{Definition}[Theorem]{Definition}
\theoremstyle{definition}
\newtheorem{Example}[Theorem]{Example}

\newtheorem{Remark}[Theorem]{Remark}

\def\fd{\mathfrak{d}}
\def\md{\boldsymbol{\fd}}
\def\ioda{\boldsymbol{\iota}}

\def\mod{\mathrm{mod}}

\def\0{{\hat{0}}}
\def\1{{\hat{1}}}

\newcommand \pd {{\partial}}

\def\HPV{\mathrm{HPV}}
\def\rH{\mathrm{H}}
\def\cO{\mathcal{O}}

\def\J{\mathrm{J}}

\def\ioda{\boldsymbol{\iota}}
\def\rGamma{\mathrm{\Gamma}}

\def\hW{{\hat W}}

\def\hOmega{{\hat \Omega}}
\def\hD{{\hat D}}

\def\an{\mathrm{an}}

\begin{document}

\title{B-type Landau-Ginzburg models on Stein manifolds}

\author{E. M. Babalic$^1$, D. Doryn$^2$, C. I. Lazaroiu$^3$
  and M. Tavakol$^4$}

\address{$^1$ Horia Hulubei National Institute for Physics and Nuclear
  Engineering (IFIN-HH), Bucharest-Magurele, Romania} \address{$^{1,2,3,4}$
  Center for Geometry and Physics, Institute for Basic Science, Pohang
  37673, Korea} \address{$^4$ School of mathematics and
  statistics, University of Melbourne, VIC 3010, Australia}

\ead{mbabalic@theory.nipne.ro, dmitry.doryn@gmail.com, calin@ibs.re.kr,
  mehdi.tavakol@unimelb.edu.au}

\begin{abstract}
We summarize the description of the open-closed TFT datum for B-type
Landau-Ginzburg models with Stein manifold targets and
discuss various constructions which lead to large classes of examples
of such models.
\end{abstract}

\section*{Introduction}

\

\noindent It is well-known \cite{LG1,LG2} that B-type classical
two-dimensional topological Landau-Ginzburg (LG) models with D-branes
can be defined for any non-compact smooth Calabi-Yau manifold target
$X$ and any holomorphic superpotential $W:X\rightarrow \C$. The
quantization of such models is expected to produce non-anomalous 2d
open-closed quantum topological field theories (TFTs), which must in turn
obey the axioms first introduced in \cite{top}. The analysis of
\cite{top} shows that such quantum field theories are equivalent with
a {\em TFT datum}, an algebraic structure subject to certain
axioms. It follows that the quantum B-type LG model with D-branes
associated to the pair $(X,W)$ is entirely encoded by its TFT datum,
which therefore should admit a description in terms of objects
naturally associated to such a pair. This description was extracted in
\cite{LG2} through path integral arguments. It was formulated
rigorously and further analyzed in \cite{lg1,lg2,tserre}.

As shown in \cite{lg2}, the Landau-Ginzburg TFT datum simplifies
dramatically when the target Calabi-Yau manifold is Stein, leading to
a large class of B-type LG models which were not considered
previously. These need not be algebraic in the sense that $X$ need not
be the analyticization of an affine complex algebraic variety and $W$
need not be the analyticization of a regular function. As a
consequence, the corresponding TFT datum is described by constructions
carried out in the analytic category.  In this contribution, we give a
brief review of B-type LG models with Stein Calabi-Yau targets,
focusing on certain constructions which lead to large classes of examples.

\section{Landau-Ginzburg pairs and B-type LG models}

\

\noindent The axiomatic treatment of \cite{top} shows that any 2d
open-closed TFT is determined by a TFT datum of parity $\mu\in \Z_2$,
which consists of:
\begin{itemize} \itemsep 0.0em
\item The on-shell {\em bulk algebra} $\cH$, which is a
finite-dimensional and $\Z_2$-graded associative and commutative
$\C$-algebra
\item The {\em topological D-brane category} $\cT$, which is a
$\Z_2$-graded and Hom-finite $\C$-linear category whose objects are
the topological D-branes of the model.
\item The {\em bulk-boundary maps} $e_a:\cH\rightarrow \Hom_\cT(a,a)$,
which are even $\C$-linear maps defined for any D-brane $a\in \Ob
\cT$.
\item The {\em boundary-bulk maps} $f_a:\Hom_\cT(a,a)\rightarrow \cH$,
which are $\C$-linear maps of $\Z_2$-degree $\mu$, defined for any
$a\in \Ob \cT$.
\item The {\em bulk trace} $\Tr:\cH\rightarrow \C$, an even and
$\C$-linear non-degenerate map having the graded trace property.
\item The {\em boundary traces} $\tr_a:\Hom_\cT(a,a)\rightarrow \C$,
which are $\C$-linear and non-degenerate maps of $\Z_2$-degree $\mu$
defined for each $a\in \Ob\cT$, such that the map $\sum_{a}\!
\tr_a\!:\oplus_{a}\Hom_\cT(a,a)\rightarrow \C$ has the graded trace
property with respect to the associative multiplication induced on
$\oplus_{a}\Hom_\cT(a,a)$ by the composition of $\cT$.
\end{itemize}

\noindent This data is subject to certain further conditions discussed in \cite{top,lg1}. 

\begin{Definition}
A {\em Landau-Ginzburg (LG) pair} is a pair $(X,W)$ such that:
\begin{enumerate}[1.]
\item $X$ is a non-compact complex and K\"ahlerian manifold of
  dimension $d>0$, which is holomorphically Calabi-Yau in the sense
  that its holomorphic canonical line bundle $K_X$ is holomorphically
  trivial.
\item $W\in \O(X)$ is a non-constant holomorphic complex-valued
  function defined on $X$.
\end{enumerate}
\end{Definition}

\noindent Given an LG pair $(X,W)$, let
$Z_W\eqdef \{x\in X| (\pd W)(x)=0\}$ denote the critical set of $W$;
this is an analytic subset of $X$, which can be highly singular in
general.

\begin{Definition}
The {\em modified contraction operator} of an LG pair $(X,W)$ is the 
sheaf morphism $\ioda_W=\i \pd W\lrcorner: TX\rightarrow \cO_X$. 
We define\footnote{We denote by $\O(X)$ the $\C$-algebra of holomorphic
complex-valued functions defined on $X$ and by $\cO_X$ the sheaf of holomorphic 
complex-valued functions defined on open subsets of $X$, while $\Gamma(X, E)$ 
denotes the $\O(X)$-module of globally-defined holomorphic sections of a holomorphic 
vector bundle $E$.}: 
\begin{itemize}
\itemsep 0.0em
\item The {\em critical sheaf} ${\cal J}_W\eqdef
  \im(\ioda_W:TX\to \cO_X)$.
\item The {\em Jacobi sheaf} $Jac_W\eqdef \cO_X/{\cal
  J}_W$.
\item The {\em Jacobi algebra} $\Jac(X,W)\eqdef
  \Gamma(X,Jac_W)$.
\item The {\em critical ideal} $\J(X,W)\eqdef
  {\cal J}_W(X)=\ioda_W(\Gamma(X,TX))\subset \O(X)$.
\end{itemize}
\end{Definition}

\noindent To any LG pair $(X,W)$, one can associate a classical B-type
open-closed topological LG model which was constructed in
\cite{LG1,LG2}. When the critical set $Z_W$ is compact, the path
integral arguments of \cite{LG2} lead to a proposal for the TFT datum
of the corresponding quantum theory, which was refined and analyzed
rigorously in \cite{lg1,lg2,tserre}; see \cite{lgproc} for a brief
review of those results. It was shown in \cite{lg1,tserre} that this
proposal satisfies all TFT datum axioms except for the topological
Cardy constraint, which is also conjectured to hold. As shown in
\cite{lg2} and summarized below, the Landau-Ginzburg TFT datum admits
a simpler equivalent description when $X$ is a Stein manifold.

\begin{Remark} The bulk algebra $\cH$ and topological D-brane category
$\cT$ of the B-type LG model parameterized by $(X,W)$ are not only
$\C$-linear but also $\O(X)$-linear (see \cite{LG1}).
\end{Remark}

\pagebreak

\section{B-type LG models with Calabi-Yau Stein manifold target}

\vspace{5mm}

\subsection{Stein manifolds}

\

\begin{Definition}
Let $X$ be a complex manifold of dimension $d>0$. We say that $X$ is
{\em Stein} if it admits a holomorphic embedding as a {\em closed}
complex submanifold of $\C^N$ for some $N>1$.
\end{Definition}

\paragraph{\bf Remarks}
\begin{itemize}
\itemsep 0.0em
\item There exist numerous equivalent definitions of Stein manifolds.
\item Any Stein manifold is K\"ahlerian.
\item The analyticization of any non-singular complex affine variety
is Stein, but most Stein manifolds are {\em not} of this type.
\end{itemize}

\noindent Since all Stein manifolds are K\"ahlerian, one can consider
B-type Landau-Ginzburg models with Stein Calabi-Yau target. 

\paragraph{\bf Examples}

\begin{itemize}
\itemsep 0.0em
\item $\C^d$ is a Stein manifold.
\item Every domain of holomorphy in $\C^d$ is a Stein manifold.
\item Every closed complex submanifold of a Stein manifold is a Stein
manifold.
\item Any non-singular analytic complete intersection in $\C^N$ is a
Stein manifold.
\item Any (non-singular) connected open Riemann surface without border
is a Stein manifold.
\end{itemize}

\noindent The following result is crucial in Stein geometry: 

\begin{Theorem}[Cartan's theorem  B]
For every coherent analytic sheaf ${\cal F}$ on a Stein manifold $X$,
the sheaf cohomology $\rH^i(X,{\cal F})$ vanishes for all $i>0$.
\end{Theorem}

\

\noindent For what follows, let $(X,W)$ to be a Landau-Ginzburg pair such that 
$X$ is a Stein Calabi-Yau manifold of complex dimension $d$.  

\subsection{The bulk algebra}

\

\begin{Theorem}{\rm \cite{LG2}}
Suppose that the critical locus $Z_W$ is compact. Then:
\begin{enumerate}[1.]
\itemsep 0.0em
\item $Z_W$ is necessarily finite.
\item The bulk algebra is concentrated in even degree and can be
  identified with the Jacobi algebra:
\be 
\cH\equiv_{\O(X)} \Jac(X,W)~~.  
\ee 
Moreover, we have an isomorphism of $\O(X)$-algebras: 
\be
\label{JacAlgebraStein}
\Jac(X,W)\simeq_{\O(X)} \O(X)/\J(X,W)~~.  
\ee
\item Suppose that $X$ is holomorphically parallelizable (i.e.  $TX$ is
  holomorphically trivial). Then:
\be 
\J(X,W)=\langle u_1(W),\ldots,  u_d(W)\rangle~~, 
\ee 
where $u_1,\ldots, u_d$ is any global holomorphic frame of $TX$ and
$\langle f_1,\ldots, f_n\rangle$ denotes the ideal of $\O(X)$
generated by the holomorphic functions $f_j\in \O(X)$.
\end{enumerate}
\end{Theorem}

\subsection{The topological D-brane category}

\

\begin{Definition} A {\em holomorphic factorization} of $W$ is a pair
$(E,D)$, where $E$ is a $\Z_2$-graded holomorphic vector bundle
defined on $X$ and $D\in \Gamma(X,\End^\1(E))$ is an odd holomorphic
section of $E$ such that $D^2=W \id_E$.
\end{Definition}

\begin{Definition} The {\em holomorphic dg category of holomorphic
factorizations} of $(X,W)$ is the $\Z_2$-graded $\O(X)$-linear dg
category $\F(X,W)$ defined as follows:
\begin{itemize}
\itemsep 0.0em
\item The objects of $\F(X,W)$ are the holomorphic factorizations of $W$.
\item The hom-sets are the $\O(X)$-modules
  $\Hom_{\F(X,W)}(a_1,a_2)\eqdef \rGamma(X,Hom(E_1,E_2))$, endowed
  with the $\Z_2$-grading:
\be
\Hom_{\F(X,W)}^\kappa(a_1,a_2)=\Gamma(X, Hom^\kappa(E_1,E_2))~,~\forall \kappa\in \Z_2
\ee
and with the {\em defect differentials} $\md_{a_1,a_2}$ determined uniquely by the
condition:
\be
\md_{a_1,a_2}(f)= D_2\circ f-(-1)^\kappa f\circ D_1~,~\forall f\in \rGamma(X, Hom^\kappa(E_1,E_2))~,~\forall \kappa\in \Z_2~~.
\ee
Here $a_1:=(E_1,D_1)$ and $a_2:=(E_2,D_2)$ are any two holomorphic factorizations of $W$.
\item The composition of morphisms is the obvious one.
\end{itemize}
Let $\HF(X,W)\eqdef \rH(\F(X,W))$ be the total cohomology category of
the dg category $\F(X,W)$.
\end{Definition}

\begin{Definition}
A {\em projective analytic factorization} of $W$ is a pair
$(P,D)$, where $P$ is a finitely generated projective
 supermodule and $D\in \End_{\O(X)}^1(P)$ is an odd
endomorphism of $P$ such that $D^2=W \id_P$.
\end{Definition}

\begin{Definition}
The {\em dg category $\PF(X,W)$ of projective analytic
  factorizations} of $W$ is the $\Z_2$-graded $\O(X)$-linear
dg category defined as follows:
\begin{itemize}
\itemsep 0.0em
\item The objects of $\PF(X,W)$ are the projective analytic factorizations of $W$.
\item The hom-sets $\Hom_{\PF(X,W)}((P_1,D_1),(P_2,D_2))\eqdef \Hom_{\O(X)}(P_1,P_2)$ are 
endowed with $\Z_2$-grading and with the odd
differential $\md:=\md_{(P_1,D_1),(P_2,D_2)}$ 
\be
\md(f)= D_2\circ f-(-1)^{\deg f}f\circ D_1f~,~~\forall f\in \Hom_{\O(X)}(P_1,P_2)~.
 \ee
\item The composition of morphisms is the obvious one.
\end{itemize}
Let $\HPF(X,W)\eqdef \rH(\PF(X,W))$ be the total cohomology category of
the dg category $\PF(X,W)$.
\end{Definition}

\noindent For any unital commutative ring $R$ and any element $r\in R$, let
$\MF(R,r)$ denote category of finite rank matrix factorizations of $W$
over $R$ and $\HMF(R,r)$ denote its total cohomology category (which
is $\Z_2$-graded and $R$-linear).

\begin{Theorem} {\rm \cite{lg2}}
Let $(X,W)$ be an LG pair such that $X$ is a Stein
manifold. Then: 
\begin{enumerate}
\itemsep 0.0em
\item There exists a natural equivalence of $\O(X)$-linear
and $\Z_2$-graded dg categories:
\be
\F(X,W)\simeq_{\O(X)} \PF(X,W)~~.
\ee
\item If the critical locus $Z_W$ is finite, then the topological
  D-brane category $\cT$ is given by:
\be
\cT\equiv \HF(X,W)\simeq_{\O(X)} \HPF(X,W)~~.
\ee
\item Multiplication with elements of the critical ideal $\J(X,W)$
acts trivially on $\HF(X,W)$, so $\cT$ can be viewed as a
$\Z_2$-graded $\Jac(X,W)$-linear category.
\item The even subcategory $\cT^\0$ has a natural
  triangulated structure.
\item There exists an $\O(X)$-linear dg functor
  $\Xi:\F(X,W)\!\rightarrow \oplus_{p\in Z_W} \!\MF(\cO_{X,p},\hW_p)$
  which induces a full and faithful $\Jac(X,W)$-linear functor
  $\Xi_\ast:\HF(X,W)\rightarrow \oplus_{p\in Z_W}
  \HMF(\cO_{X,p},\hW_p)$.
\end{enumerate}
\end{Theorem}

\subsection{The remaining objects}

\

\noindent When $X$ is a Stein manifold with complex dimension $d$, 
the remaining objects of the TFT datum are as follows (see \cite{lg2}): 
\begin{itemize}
\itemsep 0.0em
\item The {\em bulk trace} is given by\footnote{Given a holomorphic vector bundle $V$ on $X$, $\hat {\beta}_p$ denotes the germ at
    $p\in X$ of a holomorphic section $\beta\in \Gamma(X,V)$.}: 
\be
\Tr(f)=\sum_{p\in Z_W} A_p\, \Res_p\left[\frac{{\hat f}_p{\hOmega}_p}{{\det}_{{\hOmega}_p}(\pd W)}\right]~~.
\ee
\item The {\em boundary trace} of the D-brane (holomorphic factorization)
  $a=(E,D)$ is given by the sum of {\em generalized Kapustin-Li traces}\footnote{Here 
  $\str$ denotes the supertrace on the finite-dimensional $\Z_2$-graded vector space 
$\End_\C(\End_{\cT}(a))$.}:
\be
\tr_a(s)=\frac{(-1)^{\frac{d(d-1)}{2}}}{d!}\sum_{p\in Z_W} 
A_p\, \Res_p\left[\frac{\str\left({\det}_{{\hOmega}_p}(\pd {\hD}_p) {\hat s}_p\right){\hOmega}_p}
{{\det}_{{\hOmega}_p}(\pd \hW_p)}\right]~.
\ee
Here $\Omega$ is a holomorphic volume form on $X$, $A_p$ are
normalization constants and $\Res_p$ denotes the Grothendieck residue
on $\cO_{X,p}$.
\item The {\em bulk-boundary} and {\em boundary-bulk} maps of $a=(E,D)$ are given by:
\beqa
&& e_a(f)\equiv \i^d (-1)^{\frac{d(d-1)}{2}}\bigoplus_{p\in Z_W} {\hat f}_p \id_{E_p}~,~~\forall f\in \cH\equiv \Jac(X,W)\\
&& f_a(s)\equiv \frac{\i^d}{d!}\bigoplus_{p\in Z_W}\str\left({\det}_{{\hOmega}_p}(\pd {\hD}_p) \, {\hat s}_p\right)~,~~\forall s\in \End_{\cT}(a)\equiv \rGamma(X, End(E))~.
\eeqa
\end{itemize}

\subsection{Topologically non-trivial elementary holomorphic factorizations}
\label{subsec:nontriv}

\

\noindent By the Oka-Grauert principle, a holomorphic vector bundle
$V$ on a Stein manifold $X$ is holomorphically trivial iff it is
topologically trivial, in which case we simply say that it is trivial.

\begin{Definition}
Let $(X,W)$ be an LG pair where $X$ is a Stein manifold.  A
holomorphic factorization $(E,D)$ of $W$ is called {\em elementary} if
$\rk\, E^\0=\rk \,E^\1=1$ and {\em topologically trivial} if $E$ is
trivial as a $\Z_2$-graded holomorphic vector bundle, i.e. if both
$E^\0$ and $E^\1$ are trivial.
\end{Definition}

\paragraph{\bf Construction.}  Let $X$ be a Calabi-Yau Stein manifold
with $\rH^2(X,\Z)\neq 0$. Given a non-trivial holomorphic line bundle
$L$ on $X$ and non-trivial holomorphic sections $v\in \Gamma(X,L)$ and
$u\in \Gamma(X,L^{-1})$, the $\Z_2$-graded holomorphic vector bundle
$E\eqdef \cO_X\oplus L$ admits the odd global holomorphic section
$D\eqdef \left[\begin{array}{cc} 0 &v\\ u & 0\end{array}\right]$. The
tensor product $u\otimes v\in\rH^0(L\otimes L^{-1})$ identifies with a
non-trivial holomorphic function $W$ defined on $X$ through any
isomorphism $L\otimes L^{-1}\simeq \cO_X$ and $(E,D)$ is a non-trivial
elementary holomorphic factorization of $W$.

\begin{Remark} Let $\Pic_\an(X)$ denote the analytic Picard group of
holomorphic line bundles on any Stein manifold $X$. Then the
assignment $L \to c_1(L)$ induces an isomorphism $\Pic_\an (X) \simeq
\rH^2(X,\Z)$.  When $X$ is the analytic space associated to an
algebraic variety, the natural map $\Pic_\alg (X) \to \rH^2(X,\Z)$
from the algebraic Picard group need not be isomorphism.
\end{Remark}

\section{Examples}

\vspace{5mm}

\subsection{Domains of holomorphy in $\C^d$}

\

\noindent Let $X=U\subseteq \C^d$ be a domain of
holomorphy. Then $U$ is Stein and holomorphically
parallelizable (hence Calabi-Yau) with global holomorphic frame
$u_i=\partial_i$, where $\partial_i:= \frac{\partial}{\partial z^i}$
and $\{z^1,\ldots, z^d\}$ are global holomorphic coordinates on
$U$. Let $W\in \O(U)$ have finite critical set. We have \cite{lg2}:
\be
\HPV(U,W)=\HPV^0(U,W)\simeq_{\O(U)} \Jac(U,W)=\O(U)/\langle \partial_1 W,\ldots, \partial_d W\rangle~~.
\ee

\begin{Example} {\rm \cite{lg2}}
Suppose that $U$ is contractible. Then $\HDF(U,W)$ is equivalent with
the $\Z_2$-graded cohomological category $\HMF(\O(U),W)$ of analytic
matrix factorizations of $W$.
\end{Example}

\subsection{Open Riemann surfaces}

\

\noindent Let $\Sigma$ be any open Riemann surface, i.e. a smooth,
connected and non-compact complex Riemann surface without
boundary. Such a surface need not be affine algebraic; in particular,
it can have infinite genus and an infinite number of Freudenthal
ends. Any open Riemann surface is Stein and any holomorphic vector
bundle defined on it is trivial. In particular, $\Sigma$ is
holomorphically parallelizable and hence Calabi-Yau. Any
finitely-generated projective $\O(\Sigma)$-module is free, thus
$\HPF(\Sigma,W)$ coincides with the $\Z_2$-graded category
$\HMF(\O(\Sigma),W)$ of finite rank matrix factorizations of $W$ over
the ring of holomorphic functions defined on $\Sigma$.

\begin{Proposition}{\rm \cite{lg2}}
\label{Riemann}
Let $W\in \O(\Sigma)$ be a non-constant holomorphic function. Then the
bulk algebra of the corresponding B-type LG model is given by:
\be
\cH\simeq \Jac(\Sigma,W)=\O(\Sigma)/\langle v(W) \rangle~~,
\ee
where $v$ is any non-trivial globally-defined holomorphic vector field
on $\Sigma$. Moreover, there exists an equivalence of
$\O(\Sigma)$-linear $\Z_2$-graded categories:
\be
\HF(\Sigma,W)\simeq \HPF(\Sigma,W)= \HMF(\O(\Sigma),W)~~.
\ee
\end{Proposition}

\noindent More detail on B-type LG models with open
Riemann surface targets can be found in \cite{calin}.

\subsection{Analytic complete intersections}

\

\noindent Let $X\subset \C^N$ be an analytic complete intersection of
complex dimension $d$, defined by the regular sequence of holomorphic
functions $f_1, \dots , f_{N-d}\in \O(\C^N)$. Let $u_1,\ldots,u_d$ be
a global holomorphic frame of $TX$. Then $X$ is Stein and
holomorphically parallelizable (hence also Calabi-Yau).  We have
$\O(X)\simeq \O(\C^N)/\langle f_1,\ldots, f_{N-d}\rangle$. Any $W\in
\O(X)$ is the restriction to $X$ of a holomorphic function $\cW\in
\O(\C^N)$. Let $Z_\cW\subset \C^N$ be the critical locus of
$\cW$. Then $Z_\cW\cap X\subseteq Z_W$, but the inclusion may be
strict. This construction provides a large class of B-type LG models
with Stein Calabi-Yau targets. The simplest case arises when $X$ is an
analytic hypersurface in $\C^N$.

\begin{Proposition} {\rm \cite{LG2}}
Let $X$ be a non-singular analytic hypersurface in $\C^N$ defined by
the equation $f=0$. Then the holomorphic tangent vector fields
$(v_{ij})_{1\leq i<j\leq N}$ defined on $X$ through:
\be
v_{ij}^{k}=(\partial_j f)\delta_{ik} -(\partial_i f) \delta_{jk}~~~(k=1,\ldots, N)
\ee
generate each fiber of $TX$. If $\cW\in \O(\C^N)$ is a holomorphic function, then
the critical locus $Z_W$ of the restriction $W\eqdef \cW|_X$ is
defined by the system:
\ben
\label{ZWSurface}
 f=0~,~~ \partial_i \cW\partial_j f -\partial_j \cW\partial_i f=0~~~(1\leq i<j\leq N)~~.
\een
If $W$ has isolated critical points on $X$, then we have:
\ben
\label{JacSurface}
\Jac(X,W)=\O(\C^N)/I~~,
\een
where $I\subset \O(\C^N)$ is the ideal generated by $f$ and by the
holomorphic functions $\partial_i \cW\partial_j f -\partial_j
\cW\partial_i f$ with $1\leq i<j\leq N$.
\end{Proposition}

\begin{Example}
Let $X$ be the non-singular analytic hypersurface defined in $\C^3$ by
the equation $f(x_1,x_2,x_3)=x_1e^{x_2}+x_2e^{x_3}+x_3e^{x_1}=0$ and
$\cW\in \O(\C^3)$ be the holomorphic function given by
$\cW(x_1,x_2,x_3)=x_1^{n+1}+x_2x_3$, where $n\geq 1$. Let $W\eqdef
\cW|_X\in \O(X)$. The critical locus of $\cW$ coincides with the
origin of $\C^3$, which lies on $X$. Thus $Z_W$ contains
the point $(0,0,0)\in X$. In this example, we have:
\be
\pd_1 f=e^{x_2}+x_3 e^{x_1}~,~\pd_2 f=e^{x_3}+x_1 e^{x_2}~,~\pd_3 f=e^{x_1}+x_2 e^{x_3}~~.
\ee
The vector fields of Proposition 2. are given by:
\beqan
\label{vk}
&&v_{23}=(0, \pd_3 f, -\pd_2 f)=\big(0, e^{x_1}+x_2 e^{x_3}, -e^{x_3}-x_1 e^{x_2} \big)~~\nn\\
&&v_{13}=(\pd_3 f, 0, -\pd_1 f)=\big(e^{x_1}+x_2 e^{x_3}, 0, -e^{x_2}-x_3 e^{x_1}\big)~~\\
&&v_{12}=(\pd_2 f, -\pd_1 f ,0)=\big(e^{x_3}+x_1 e^{x_2}, -e^{x_2}-x_3 e^{x_1},0 \big)~~\nn
\eeqan
and the defining equations \eqref{ZWSurface} of $Z_W$ take the form:
\beqan
\label{Zeq1}
& x_1 e^{x_2}+x_2 e^{x_3} +x_3 e^{x_1} =0~~\nn\\
& (n+1)x_1^n(e^{x_1}+x_2 e^{x_3})-x_2 (e^{x_2}+ x_3 e^{x_1})=0~~\nn\\
& x_3 (e^{x_1}+x_2 e^{x_3}) - x_2 (e^{x_3}+x_1 e^{x_2})=0~~\\
& (n+1) x_1^n (e^{x_3} +  x_1e^{x_2}) - x_3 (e^{x_2} + x_3 e^{x_1} )=0~~.\nn
\eeqan
The bulk space $\cH$ is the Jacobi algebra $\Jac(X,W)=\O(\C^3)/I$,
where $I\subset \O(\C^3)$ is the ideal generated by $f$ and the four
holomorphic functions appearing in the left hand side of the previous
system. Numerical study shows that, for generic $n\geq 1$, the above
transcendental system admits solutions different from $x_1=x_2=x_3=0$,
so $W$ has critical points on $X$ which differ from the origin. For
any $k\in\{0,\ldots, n+1\}$, an example of holomorphic (here even
algebraic) factorization $(E,D_k)$ of $\cW$ on $\C^3$ is obtained by
taking $E^\0=E^\1=\cO_{\C^3}^{\oplus 2}$ with:
\be
D_k=\left[\begin{array}{cc} 
0 & b_k \\
a_k & 0
\end{array}\right]~~,~~\mathrm{where}~~ a_k=\left[\begin{array}{cc} 
x_2 &~ x_1^{n+1-k} \\
x_1^k & -x_3
\end{array}\right]~~\mathrm{and}~~b_k=\left[\begin{array}{cc} 
x_3 &~ x_1^{n+1-k} \\
x_1^k & -x_2
\end{array}\right]~~
\ee
This induces a holomorphic factorization $(E|_X,D_k|_X)$ of
$W$.

\begin{figure}
\begin{center}
\includegraphics[height=3cm]{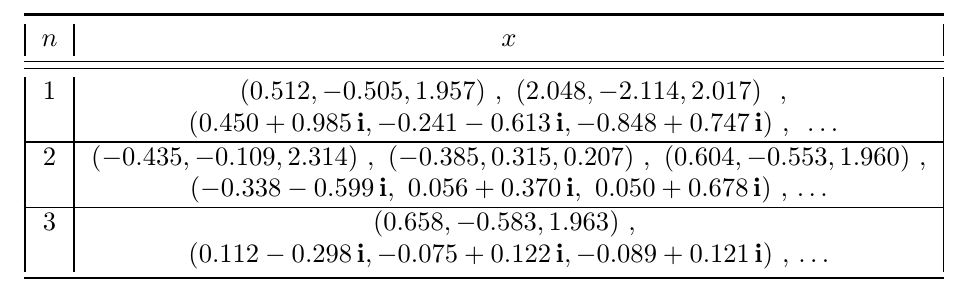}
 \caption{Some non-zero critical points of $W$ on $X$
 for small $n$ (4 significant digits).}
\end{center}
\end{figure}
\end{Example}

\subsection{Complements of affine hyperplane arrangements}

\

\noindent Let $\cA$ be a $d$-dimensional central complex affine
hyperplane arrangement, i.e. a finite set of distinct linear
hyperplanes $H=\ker \alpha_H\subset \C^d$, where
$\alpha_H\!:\C^d\rightarrow \C$ are non-trivial linear functionals. The
complex manifold $X\!\eqdef\! \C^d \setminus \left[\cup_{H\in
\cA}H\right]$ is Stein and parallelizable (hence Calabi-Yau)
and coincides with the analyticization of the affine hypersurface in
$\C^{d+1}$ defined by:
\be
x_{d+1}\prod_{H\in \cA}{\alpha_H(x_1,\ldots, x_d)}=1~~.
\ee
The cohomology ring $\rH(X,\Z)$ is
isomorphic as a graded $\Z$-algebra with the {\em Orlik-Solomon
algebra} of $\cA$, whose homogeneous components $A^k$ are free
$\Z$-modules of finite rank. In particular, the cohomology groups
$\rH^k(X,\Z)$ are free Abelian groups of finite rank. The {\em
intersection poset $L$} of $\cA$ is the set of all those subspaces
$F\subset \C^d$ (called {\em flats}) which arise as finite
intersections of hyperplanes from $\cA$, ordered by reverse
inclusion. Since $\cA$ is central, its intersection poset $L$ is a
bounded geometric lattice with greatest element given by $\cap_{H\in
\cA}H$. Let $P_X(t)\eqdef \sum_{j=0}^d \rk \,\rH^j(X,\Z) \, t^j$ be
the Poincar\'e polynomial of $X$.

\begin{Proposition}{\rm \cite[Theorem 2.2 \& Corollary
    3.6]{Dimca}}
Let $\mu_L$ be the M\"{o}bius function of the locally-finite poset
$L$. Then $\mu_L(\C^d,F)\neq 0$ for all flats $F\in L$ and the sign of
$\mu_L(\C^d,F)$ equals $(-1)^{\codim_\C F}$. Moreover, we have:
\be
P_X(t)=\sum_{F \in L} |\mu_L(\C^d,F)| t^{\codim_\C F}=\sum_{F \in L} \mu_L(\C^d,F) (-t)^{\codim_\C F}~~.
\ee
\end{Proposition}

\noindent One has $\rk \,\rH^j(X,\Z)>0$ for all $j=0,\ldots, \rk \,\cA$
and $\rk \,\rH^j(X,\Z)=0$ for $j>\rk\, \cA$ (see \cite[Chap. 2.5, Exercise
2.5]{Dimca}), where $\rk\,\cA\eqdef \codim_\C \left[\cap_{H\in \cA}
H\right]$ is the {\em rank} of $\cA$. This implies \cite{lg2} that the
complement of any $d$-dimensional central complex affine hyperplane
arrangement $\cA$ with $d\geq 2$ and $\rk\,\cA\geq 2$ admits
topologically non-trivial elementary holomorphic factorizations.

\begin{Example}
Let $\cA\subset \C^3$ be the 6-hyperplane arrangement defined by the
linear functionals $x,y,z,x-y,x-z$ and $y-z$. Then 
$P_X(t)=1+6 t+11 t^2+6 t^3$ and hence $\rH^2(X,\Z)\simeq \Z^{11}$.
\end{Example}

\begin{Example}
For any $d\geq 2$, let $\cA\subset \C^d$ be the {\em Boolean
arrangement}, defined by the equation $x_1\cdot\ldots \cdot x_d=0$.
Then $X\eqdef \C^d\setminus \left(\cup_{H\in \cA}H\right)
=(\C^\ast)^d$ is a complex algebraic torus which can be identified
with the analytic space associated to the complex affine variety:
\be
X_\alg\eqdef \Spec (\C[x_1,\ldots, x_d,x_1^{-1},\ldots, x_d^{-1}])=
\Spec\Big(\C[x_1,\ldots, x_{d+1}]/\langle x_1\cdot\ldots \cdot x_{d+1}-1\rangle \Big)~~.
\ee
In this case, the Orlik-Solomon
algebra coincides with the Grassmann $\Z$-algebra on $d$
generators and the Poincar\'e polynomial of $X$ is $P_X(t)=(t+1)^d$
(see \cite[Example 2.12 \& Example 3.3]{Dimca}), hence
$\rH^2(X,\Z)\simeq \Z^{\frac{d(d-1)}{2}}$.  One can show \cite{lg2}
that $\Pic_\alg(X)=0$ while $\Pic_\an(X)\simeq
\Z^{\frac{d(d-1)}{2}}$. Thus $X$ supports topologically non-trivial
elementary holomorphic factorizations, even though all algebraic
elementary factorizations on $X$ are topologically trivial.
\end{Example}

\begin{Example}
Let $X\simeq (\C^\ast)^2$ be the complement of the $2$-dimensional
Boolean hyperplane arrangement, which embeds in $\C^3$ as the affine
hypersurface with equation $x_1 x_2 x_3=1$. We have $\Pic_\alg(X)=0$
but $\Pic_\an(X)\simeq \rH^2(X,\Z)\simeq \Z$.  To describe
$\Pic_\an(X)$ explicitly, write $X=\C^2/\Z^2$, where $\Z^2$ acts on
$\C^2$ by:
\be
(n_1,n_2)\cdot (z_1,z_2)\eqdef (z_1+n_1,z_2+n_2)~~,
~~\forall (z_1,z_2)\in \C^2~,~\forall (n_1,n_2)\in \Z^2~~.
\ee
Consider the lattice $\Lambda:=\Z\oplus \i \Z\subset \C$ and the
elliptic curve $X_0\eqdef \C/\Lambda$ of modulus
$\tau=\i$. The maps $\varphi^\pm:\C^2\rightarrow \C$ given by
$\varphi^\pm(z_1,z_2)=z_1\pm \i z_2$ induce surjections
$\bvarphi^\pm:X\rightarrow X_0$ which are homotopy retractions of $X$
onto $X_0$. Pullback by $\bvarphi^\pm$ induces isomorphisms:
\be
\bvarphi^\pm:\NS(X_0)\rightarrow \rH^2(X,\Z)~~,
\ee
which differ only by sign, where $\NS(X_0)\eqdef
\Pic_\an(X_0)/\Pic_\an^0(X_0)=\Pic_\alg(X_0)/\Pic^0_\alg(X_0)\simeq
\rH^2(X_0,\Z)$ is the Neron-Severi group\footnote{Notice that
$\Pic_\an(X_0)=\Pic_\alg(X_0)$ by the GAGA correspondence since $X_0$
is a projective variety.} of $X$. Let $p_0\in X_0$ be the $\mod\,
\Lambda$ image of the point $z_0=\frac{1+\i}{2}\in \C$ and consider
the holomorphic line bundle $\cL=\cO_{X_0}(p_0)$ on $X_0$. Then the
class of $\cL$ modulo $\Pic^0_\an(X_0)$ generates $\NS(X_0)$ and it
was shown in \cite{lg2} that the pullbacks
$L_\pm:=(\bvarphi^\pm)^\ast(\cL)$ are holomorphic line bundles defined
on $X$, each of which generates $\Pic_\an(X)$ and which satisfy
$L_-=L_+^{-1}$. Up to multiplication by a non-zero complex number,
there exists a unique holomorphic section of $\cO_{X_0}(p_0)$ which
vanishes at $p_0$. A convenient choice $s_0\in \Gamma(\cO_{X_0}(p_0))$
is described by the Riemann-Jacobi theta function (traditionally
denoted by $\vartheta_{00}$ or $\vartheta_3$) at modulus $\tau=\i$:
\be
\vartheta(z)=\vartheta_{00}(z)|_{\tau=\i}=\sum_{n\in \Z}e^{-\pi n^2+2\pi \i n z}~~,
\ee
which satisfies: 
\be
\vartheta(z+1)=\vartheta(z)~~,~~\vartheta(z\pm \i)=e^{\pi \mp 2\pi \i z} \vartheta(z)~~
\ee
and vanishes on the lattice $\frac{1+\i}{2}+\Lambda$. The
$\bvarphi^\pm$-pullbacks of $s_0$ give global holomorphic sections
$s_\pm\in \rH^0(L_\pm)$, which are described by the
$\Z^2$-quasiperiodic holomorphic functions $f_\pm\in \O(\C^2)$ defined
through:
\be
f_\pm(z_1,z_2)\eqdef \vartheta(z_1\pm \i z_2)=\sum_{n\in \Z}e^{-\pi n^2+2\pi \i n (z_1\pm \i z_2)}~~.
\ee
The tensor product $s_+\otimes s_-\in \rH^0(L_+\otimes L_-)\simeq \O(X)$
corresponds to the holomorphic function $f\eqdef f_+f_- \in \O(\C^2)$,
which satisfies:
\be
f(z_1+1,z_2)=f(z_1,z_2)~~,~~f(z_1,z_2+1)=e^{2\pi +4\pi z_2} f(z_1,z_2)~~.
\ee
The isomorphism $L_+\otimes L_-\simeq \cO_X$ is realized on
$\Z^2$-factors of automorphy by the holomorphic
function $S:\C^2\rightarrow \C^\ast$ given by:
\be
S(z_1,z_2)\eqdef e^{-2\pi z_2^2}~~,
\ee
which satisfies $\frac{S(z_1,z_2)}{S(z_1,z_2+1)}=e^{2\pi +4\pi
  z_2}$. The section $s_+\otimes s_-\in \rH^0(L_+\otimes L_-)$
corresponds through this isomorphism to a holomorphic function $W\in
\rH^0(\cO_X)=\O(X)$ whose lift to $\C^2$ is the
$\Z^2$-periodic function:
\be
\widetilde{W}(z_1,z_2)\eqdef S(z_1,z_2)f(z_1,z_2)=e^{-2\pi z_2^2} \vartheta(z_1+\i z_2)\vartheta(z_1-\i z_2)~~.
\ee
Applying the construction of Subsection \ref{subsec:nontriv} gives a topologically
non-trivial elementary holomorphic factorization $(E,D)$ of $W$, where
$E=\cO_X\oplus L_+$ and $D=\left[\begin{array}{cc} 0 & s_-\\ s_+
    & 0\end{array} \right]$.
\end{Example}

\subsection{Complements of anticanonical divisors in Fano manifolds} 
\label{subsec:complements}

\

\noindent The following result produces a large class of Calabi-Yau
Stein manifolds which are analytifications of non-singular complex affine
varieties:

\begin{Proposition}{\rm \cite{lg2}}
\label{CY} Let $Y$ be a non-singular complex projective Fano variety
and $D$ be a smooth anticanonical divisor on $Y$. Then the
analytification of the complement $X:=Y \setminus D$ is a non-compact
Stein Calabi-Yau manifold.
\end{Proposition}

\begin{Example}
Let $\P^d$ be the $d$-dimensional projective space
with $d\geq 2$.  In this case, we have $K_{\P^d}=\O_{\P^d}(-d-1)$. Any
irreducible smooth hypersurface $Z$ of degree $d+1$ in $\P^d$ defines
an anticanonical divisor, whose complement $X=\P^d \setminus Z$ is a
Stein Calabi-Yau manifold. In this example, we have
$\rH_1(X,\Z)=\Z_{d+1}$ and the torsion part of $\rH^2(X,\Z)$ is
isomorphic with $\Ext^1(\rH_1(X,\Z),\Z)=\Ext^1(\Z_{d+1},\Z) \simeq
\Z_{d+1}$. Thus $X$ admits non-trivial holomorphic line bundles and
hence also topologically non-trivial elementary holomorphic
factorizations.
\end{Example}

\subsection{The total space of a holomorphic vector bundle}
\label{subsec:totspace}

\

\noindent Let $Y$ be a Stein manifold and $\pi: E \rightarrow Y$ be a
holomorphic vector bundle such that $c_1(E)=c_1(K_Y)$. Then the total
space $X$ of $E$ is Stein and Calabi-Yau. A particular case of this
construction is obtained by taking $E=K_Y$ of $Y$.

\begin{Example}{\rm \cite{lg2}} Consider the hypersurface $Z=\{[x,y,z]
\in \P^2 \, | \, x^2+y^2+z^2=0 \}$ in $\P^2$. Then $Y\eqdef
\P^2\setminus Z$ is Stein with $\rH^2(Y,\Z)\simeq \Z_2$ and
$c_1(K_Y)=-\gamma$, where $\gamma$ is the generator of $\rH^2(Y,\Z)$.
Let $X$ be the total space of of $K_Y$. The pullback $L$ of $K_Y$ to
$X$ has non-trivial first Chern class, hence the $\Z_2$-graded
holomorphic vector bundle $E=\cO_X\oplus L$ supports holomorphic
factorizations of some non-zero function $W\in \O(X)$.
\end{Example}

\ack{

\

\noindent This work was supported by the research grants IBS-R003-S1
and IBS-R003-D1. The work of E.M.B. was also supported by the joint
Romanian-LIT, JINR, Dubna Research Project, theme
no. 05-6-1119-2014/2019 and by the Romanian government grant PN
18090101/2019.}

\section*{References}

\

\end{document}